\documentclass[12pt,preprint,preprintnumbers,amsmath,amssymb]{revtex4}
\usepackage{graphicx}% Include figure files
\usepackage{pdfpages}
\usepackage{epsfig,natbib,color}
\usepackage{lscape}
\usepackage{graphicx}
\usepackage{epsfig}
\usepackage{natbib}

\newcommand{\aap}{    {\it Astron. Astrophys.}}
\newcommand{\apjl}{   {\it Astrophys. J. Lett.}}
\newcommand{\apjj}{   {\it Astrophys. J.}}
\newcommand{\aapr}{   {\it Astron. Astrophys. Rev.}}

\newcommand{\jgrs}{   {\it J. Geophys. Res. (Space Phys.)}}
\newcommand{\mnras}{  {\it Mon. Not. Roy. Astron. Soc.}}
\newcommand{\nn}{   {\it Nature}}
\newcommand{\natp}{   {\it Nature Phys.}}
\newcommand{\natc}{   {\it Nature Comm.}}
\newcommand{\php}{ {\it Phys. Plasmas}}
\newcommand{\solphys} { {\it Solar Phys.}}
\newcommand{\ssr}{    {\it Space Sci. Rev.}}
\newcommand{\science}{   {\it Science}}
\newcommand{\remop}{   {\it Rev. Mod. Phys.}}

\newcommand{\rslptsa}{   {\it Phil. Trans. R. Soc. A}}
\newcommand{\prll}{   {\it Phys. Rev. Lett.}}

\begin{document}
\title{Extreme Ultraviolet Imaging of Three-dimensional Magnetic Reconnection in a Solar Eruption}

\author{J. Q. Sun$^{1}$, X. Cheng$^{1\dag}$, M. D. Ding$^{1\dag}$, Y. Guo$^{1}$, E. R. Priest$^{2}$, C. E. Parnell$^{2}$, S. J. Edwards$^{3}$, J. Zhang$^{4}$, P. F. Chen$^{1}$, C. Fang$^{1}$\\}
 
\affiliation{$^{1}$School of Astronomy and Space Science, Nanjing University, Nanjing 210093, China\\}
\affiliation{$^{2}$School of Mathematics and Statistics, University of St. Andrews, Fife, KY16 9SS, Scotland, UK\\}
\affiliation{$^{3}$Department of Mathematical Sciences, Durham University, Durham DH1 3LE, UK\\}
\affiliation{$^{4}$School of Physics, Astronomy and Computational Sciences, George Mason University, Fairfax, VA 22030, USA\\}

\maketitle

\textbf{Magnetic reconnection, a change of magnetic field connectivity, is a fundamental physical process in which magnetic energy is released explosively. It is responsible for various eruptive phenomena in the universe. However, this process is difficult to observe directly. Here, the magnetic topology associated with a solar reconnection event is studied in three dimensions (3D) using the combined perspectives of two spacecraft. The sequence of extreme ultraviolet (EUV) images clearly shows that two groups of oppositely directed and non-coplanar magnetic loops gradually approach each other, forming a separator or quasi-separator and then reconnecting. The plasma near the reconnection site is subsequently heated from $\sim$1 to $\ge$5 MK. Shortly afterwards, warm flare loops ($\sim$3 MK) appear underneath the hot plasma. Other observational signatures of reconnection, including plasma inflows and downflows, are unambiguously revealed and quantitatively measured. These observations provide direct evidence of magnetic reconnection in a 3D configuration and reveal its origin.}

%(Solar Terrestrial Relations Observatory and Solar Dynamics Observatory)
%Other observational signatures of reconnection, including plasma inflows and downflows, are unambiguously revealed and quantitatively measured

Magnetic reconnection plays an important role in various astrophysical, space, and laboratory environments\cite{priest00} such as $\gamma$-ray bursts\cite{dai06}, accretion disks\cite{balbus98,yuan09}, solar and stellar coronae\cite{sturrock66,shibata07}, planetary magnetospheres\cite{phan06,xiao07}, and plasma fusion\cite{yamada10,zhong10}. In the classic two-dimensional (2D) model, reconnection occurs at an X-point where anti-parallel magnetic field lines converge and interact. As a consequence, free energy stored in the magnetic field is rapidly released and converted into other forms of energy, resulting in heating and bulk motions of plasma and acceleration of non-thermal particles\cite{priest14}. In the past decades, much attention has been paid to validate this picture. One piece of direct evidence is from in situ solar wind measurements at the magnetosheath and magnetotail of the Earth\cite{paschmann79}. Most observational evidence is from remote sensing observations of solar flares, including cusp-shaped flare loops\cite{masuda94}, plasma inflows/outflows\cite{yokoyama01,li09}, downflows above flare arcades\cite{mckenzie09}, double hard X-ray coronal sources\cite{sui03}, current sheets\cite{lin05}, and changes in connectivity of two sets of EUV loops during a compact flare\cite{su13,yang15}. With many of these observations, researchers were trying to reveal the 2D aspects of reconnection. However, reconnection is in reality a process in 3D that occurs in places where magnetic connectivity changes significantly, namely, at null points\cite{priest09}, separators\cite{longcope05,parnell10} or quasi-separators\cite{aulanier05}. 

Recently launched spacecraft Solar Terrestrial Relations Observatory (STEREO) and Solar Dynamics Observatory (SDO) provide us an unprecedented opportunity to observe reconnection in a 3D setting. Utilizing stereoscopic observations from these two spacecraft, 3D configurations of various solar phenomena have been reconstructed\cite{byrne10,feng12}. Here, we study reconnection through its reconstructed 3D magnetic topology as well as many other signatures. The Extreme Ultraviolet Imager (EUVI)\cite{howard08} on board STEREO and the Atmospheric Imaging Assembly (AIA)\cite{lemen12} on board SDO provide the necessary observational data; in particular, the AIA has an unprecedented high spatial resolution (0.6 arcsec per pixel), high cadence (12 seconds), and multi-temperature imaging ability (10 passbands).

\textbf{Results}

\begin{figure*} %%%%%%%%%%%%%%%%%% FIGURE 1
      \vspace{-0.0\textwidth}    % Shift back to the panel bottom
      \centerline{\hspace*{0.00\textwidth}
      \includegraphics[width=0.8\textwidth,clip=]{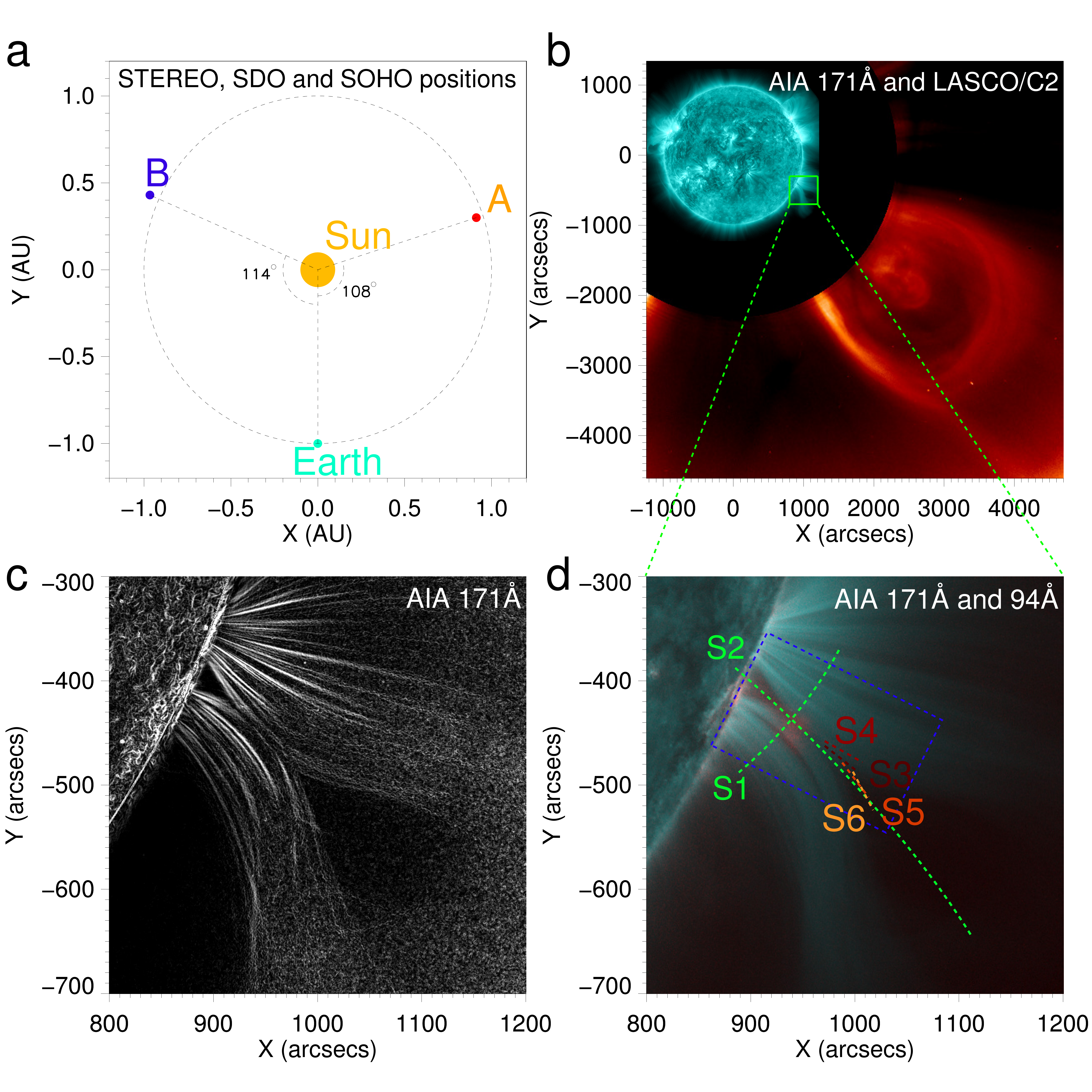}
      }
\caption{\textbf{Overview of the 2012 January 17 solar flare and CME reconnection event.} (a) The positions of the Sun, Earth, and STEREO-A/B satellites (SOHO is at L1 point and SDO is in the Earth orbit). (b) A composition of the AIA 171 {\AA} passband image (cyan) and the LASCO C2 white-light image (red). The green box indicates the main flare region. (c) The enhanced AIA 171 {\AA} image showing a clear X-shaped structure. (d) A composite image of the AIA 171 {\AA } (cyan) and 94 {\AA} (red) passbands. Cyan (red) indicates coronal loops with a temperature of $\sim$0.6 MK ($\sim$7.0 MK). Six dashed lines denote six slices (S1--S6) that are used to trace the evolution of various reconnection features with time.} \label{f1}
\end{figure*}

\textbf{Overview of the reconnection event.} The event of interest occurred on 2012 January 27, when STEREO-A and SDO were separated in space by 108 degrees along their ecliptic orbits (Figure \ref{f1}a). From $\sim$00:00 to 03:00 UT (Universal Time), a pre-existing large-scale cavity, which refers to the dark region in the EUV or soft X-ray passbands and is usually interpreted to be the cross section of a helical magnetic flux rope \cite{gibson10,cheng11,zhang12}, appears above the western solar limb as seen from the Earth. The reason why the cavity is dark may be that the density has decreased or that the plasma temperature has increased to a value outside the effective response of the lower temperature passbands. The cavity, mostly visible in the AIA 171 {\AA} passband (sensitive to a plasma temperature of $\sim$0.6 MK), starts to expand and rise from $\sim$01:40 UT, and finally results in a coronal mass ejection (CME) that is well observed by the AIA and the Large Angle and Spectrometric Coronagraph (LASCO)\citep{bru95} on board Solar and Heliospheric Observatory (SOHO; Figure \ref{f1}b). The slow rise of the cavity causes its two legs, which are made of cool loops, to approach each other and form an X-shaped structure near 03:00 UT (Figure \ref{f1}c, \ref{f1}d, and Supplementary Movie 1 and Movie 2). Following the disappearance of the cool loops (cyan in Figure \ref{f1}d), a hot region ($\sim$7 MK; visible in the AIA 94 {\AA} passband) immediately appears near the X-shaped structure, indicating the initial heating of a solar flare (red in Figure \ref{f1}d). Unfortunately, since the flare soft X-ray emission is very weak and submerged in the emission from the decay phase of a previous flare, the accurate magnitude of the flare is not recorded by Geostationary Operational Environmental Satellite. We also note that there are no X-ray observations from the Reuven Ramaty High Energy Solar Spectroscopic Imager because of annealing.

\begin{figure*} %%%%%FIGURE 2%%%%%%%%%%%
\vspace{-0.0\textwidth}    % Shift back to the panel bottom
     \centerline{\hspace*{-0.06\textwidth}
               \includegraphics[width=0.8\textwidth,clip=]{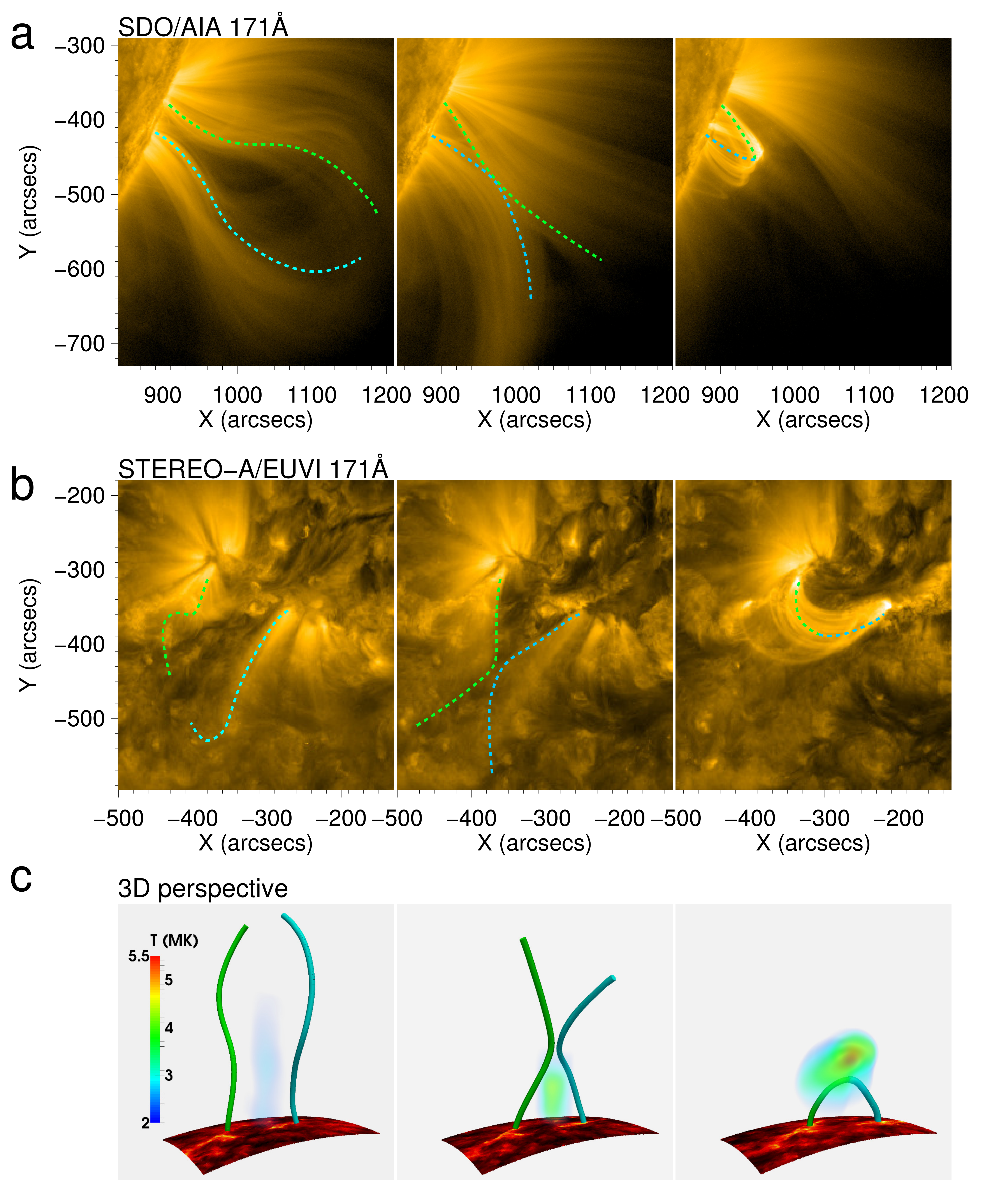}
               }
\caption{\textbf{Plasma and magnetic configurations during the reconnection process.} (a) The AIA 171 {\AA} images at 02:14 UT (left), 04:14 UT (middle), and 08:14 UT (right) displaying the side view of the evolution of two sets of coronal loops. The cyan and green dashed curves show selected coronal loops representing two magnetic field lines involved in the process. (b) The EUVI 171 {\AA} images showing the top view of the reconnection. The cyan and green dashed curves give another view of the same loops as in panel a. (c) The reconstructed 3D magnetic topology (cyan and green curves) and heated regions (cloud-like structures) before, during, and after the reconnection. The bottom boundaries are the projected EUVI 304 {\AA} images showing the footpoints of the flare and the separation of two flare ribbons.} \label{f2}
\end{figure*}

\textbf{3D topology and origin of magnetic reconnection.} Observations from SDO (the AIA 171 {\AA} passband; Figure \ref{f2}a), in combination with STEREO-A observations (the EUVI 171 {\AA} passband; Figure \ref{f2}b), enable us to reconstruct the 3D topology of the reconnection and its evolution. Due to the high magnetic Reynolds number of the ionized corona, the plasma is frozen to the magnetic field; and so the loop-like plasma emission is reasonably assumed to outline the geometry of the magnetic field\cite{priest02}. We select two magnetic loops (cyan and green dashed lines in Figure \ref{f2}a and \ref{f2}b) that can most clearly exhibit the reconnection process. With images from two perspective angles, the 3D structure of the loops is reconstructed (Figure \ref{f2}c and Supplementary Movie 3). The results display a clear picture of how the connectivity of the loops changes as the reconnection proceeds. Before reconnection, two nearly oppositely directed loops are anchored respectively at each side of the filament in the active region (left panel of Figure \ref{f2}b). The plasma between their legs has been heated to a moderate temperature (left panel of Figure \ref{f2}c). 

With the rise of the cavity, the underlying loops of opposite polarities gradually approach each other. Since the inward movements of the loops are not coplanar, an apparent separator or quasi-separator appears at $\sim$04:14 UT (middle panel of Figure \ref{f2}b). We calculate the 3D global magnetic field on January 26 using the potential field assumption\cite{sakurai89} and find an absence of pre-existing null points and separators in the reconnection region. However, the simple magnetic field in the original bipolar source region is strongly sheared from January 21 as shown by the long-existing filament/prominence at the bottom of the cavity (Figure \ref{f2}b). It suggests that a new separator or quasi-separator is formed with the prominence taking off (middle panel of Figure \ref{f2}c). As the reconnection initiates, free magnetic energy starts to be released, the most obvious consequence of which is to form a hotter region underneath the reconnection site. 

Topologically, the reconnection between the two groups of loops forms poloidal field lines above the reconnection site, increasing the twist of the erupted flux rope. At the same time, a cusp-shaped field below the reconnection site quickly shrinks into a semicircular shape to form flare loops\cite{forbes96} (right panel of Figure \ref{f2}c). With the acceleration of the CME, more plasma is heated to temperatures up to $\sim$5 MK, suggesting an enhanced reconnection. However, the heated region is still confined between the reconnection site and the flare loop top but with a spatial extension. 

\begin{figure*} %%%%%%%%%%%%%%%%% FIGURE 3
     \vspace{-0.0\textwidth}    % Shift back to the panel bottom
     \centerline{\hspace*{-0.00\textwidth}
               \includegraphics[width=0.9\textwidth,clip=]{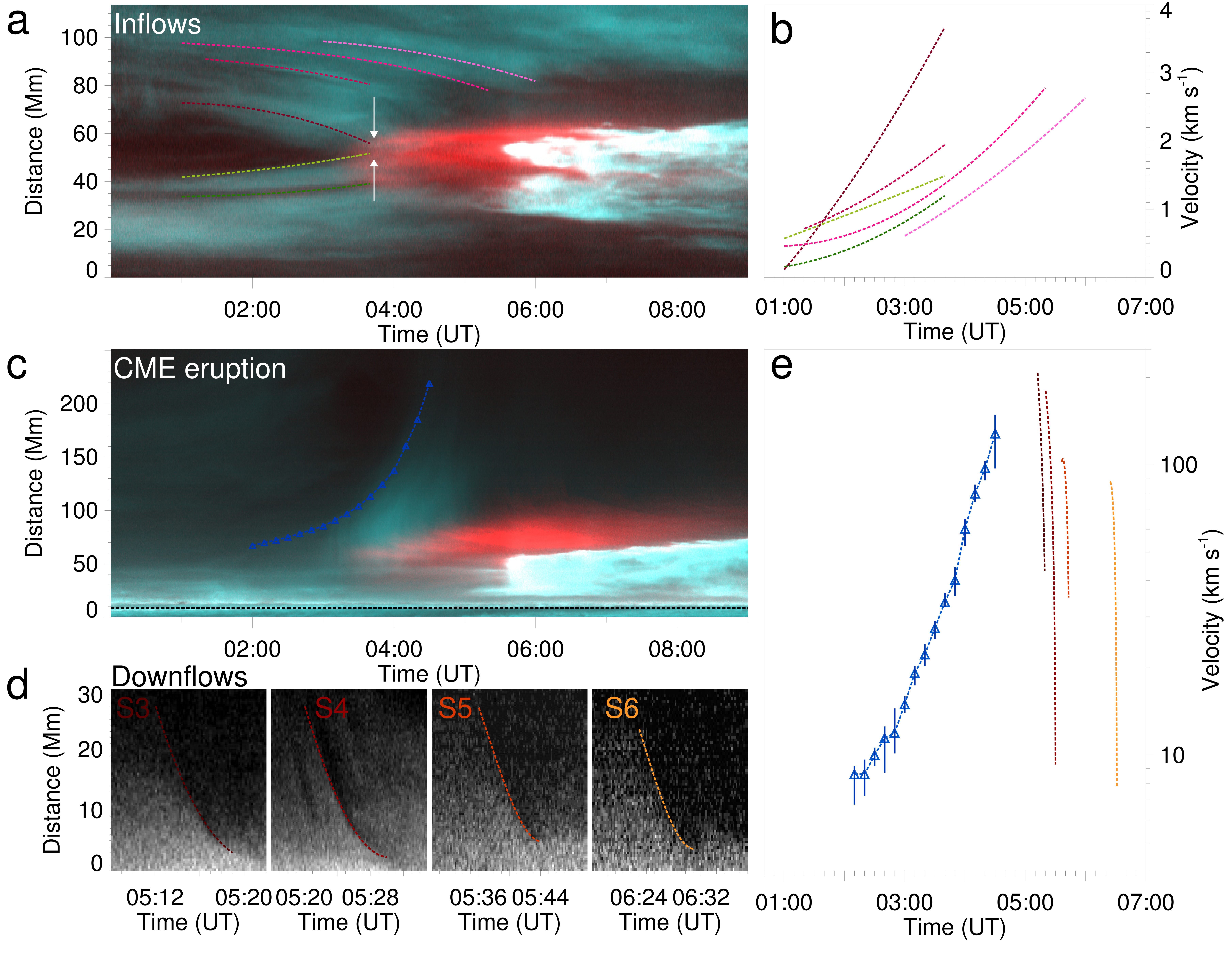}
               }
\caption{\textbf{Temporal evolution of plasma inflows and downflows during the reconnection.} (a) A time-distance plot of the composite AIA 171 {\AA} (cyan) and 94 {\AA} (red) images along the direction of the inflows (denoted by S1 in Figure \ref{f1}d) showing the approach of oppositely directed loops (two white arrows). The dashed lines with different colors (green to pink) denote the height-time measurements of the inflow at different locations. (b) The velocities of inflows, which are derived by cubic-fitting to the height-time data. (c) A time-distance plot of the composite AIA 171 {\AA} (cyan) and 94 {\AA} (red) images along the rising direction of the CME (denoted by S2 in Figure \ref{f1}d). The blue dashed line denotes the height-time measurement of the CME bubble. (d) The time-distance plots of the AIA 94 {\AA} images along the direction of four selected downflows (S3--S6 in Figure \ref{f1}d). The dashed lines are the height-time measurements of the downflows. (e) The velocities of the CME (blue) and four downflows (brown to yellow).}\label{f3}
\end{figure*}

\textbf{Quantitative properties of magnetic reconnection.}
AIA observations with high spatial and temporal resolution successfully capture evidence for reconnection including bilateral inflows, instantaneous heating of plasma near the reconnection site, and downflows that are related to the reconnection outflows\cite{mckenzie09}. To quantitatively investigate the inflows, we select an oblique slice in the 171 and 94 {\AA} composite images (S1 in Figure \ref{f1}d). The time-distance plot (Figure \ref{f3}a) clearly shows that the bilateral cool loops (cyan) keep converging to the middle (reconnection) region from $\sim$00:00 UT. Once the visible innermost loops come into contact at $\sim$03:10 UT, they immediately disappear; meanwhile, hot plasma (red) appears at the reconnection site. Several trajectories for inflows are tracked. The velocities of the inflows vary from 0.1 to 3.7 km s$^{-1}$ (Figure \ref{f3}b). Moreover, for each trajectory, the velocity tends to increase towards the reconnection site, indicating that an inward force exists on both sides of the current sheet to accelerate the inflows. 

The time-distance plot for the vertical slice (S2) shows the eruption of the CME and the downflows above the flare loops (Figure \ref{f3}c). In the early phase, only the slow rise of the CME cavity is detectable. However, along with the fast eruption of the CME, magnetic reconnection is initiated, which causes the plasma at the bottom to be rapidly heated. During this process, many dark voids intermittently appear above the heated region, rapidly falling, propagating a distance of $\sim$20--30 Mm, and finally disappearing. The time-distance plots (Figure \ref{f3}d) for four selected slices (S3--S6) show that the different downflows have almost similar trajectories. Their velocities range from $\sim$100 to 200 km s$^{-1}$ initially, but quickly decrease to tens of km s$^{-1}$ (Figure \ref{f3}e).

\begin{figure*} %%%%%%%%%%%%%%%%% FIGURE 4
     \vspace{-0.0\textwidth}    % Shift back to the panel bottom
     \centerline{\hspace*{0.00\textwidth}
               \includegraphics[width=0.8\textwidth,clip=]{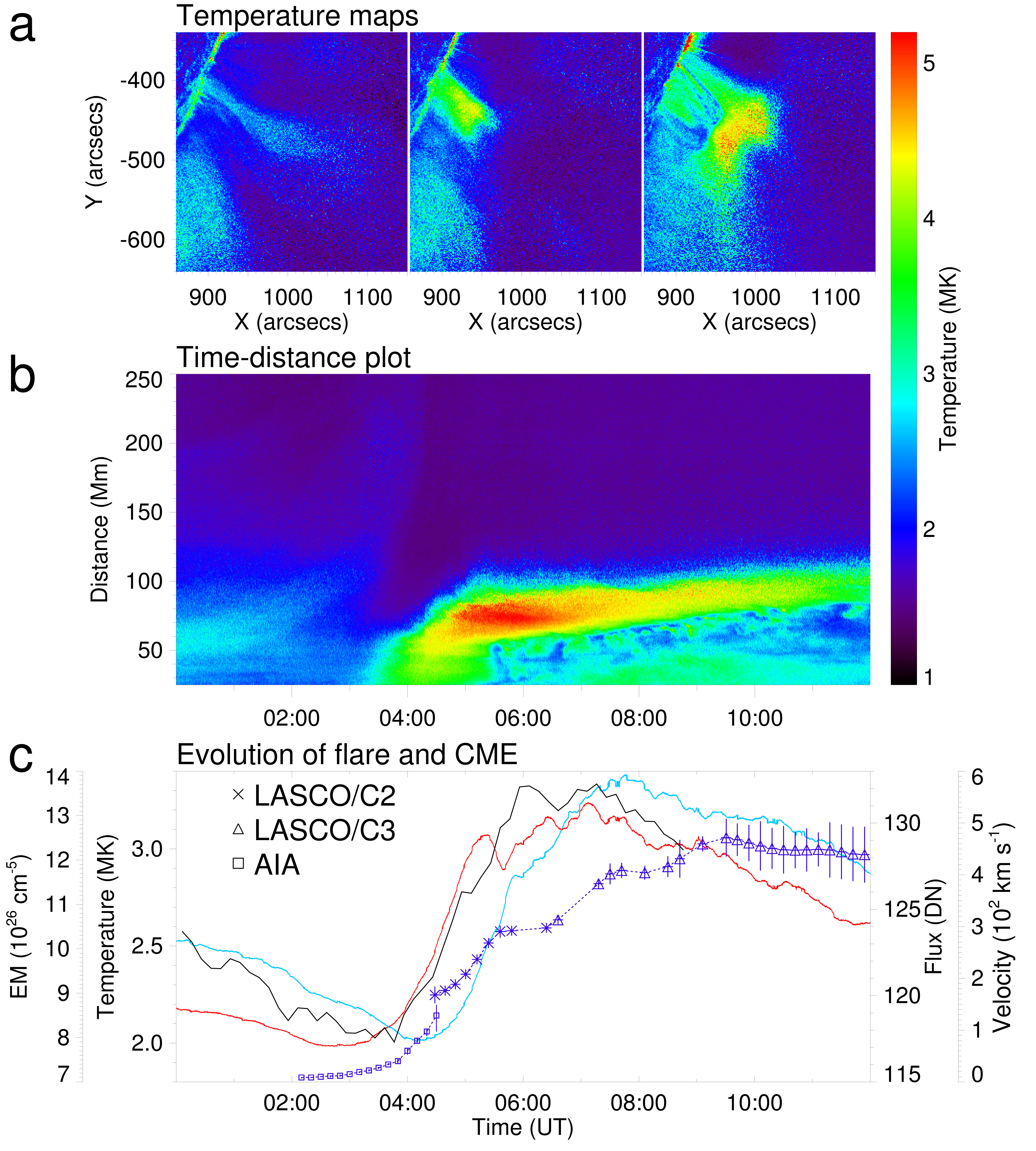}
               }
\caption{\textbf{Temporal evolution of the flare heating and the CME acceleration.} (a) The DEM-weighted temperature map at three instants (02:14, 04:14, and 08:14 UT) showing the location of the region heated by the reconnection. (b) A time-distance plot of the temperature map along the rising direction of the CME (denoted by S2 in Figure \ref{f1}d) illustrating the temperature evolution of the CME bubble and the flare region. (c) The temporal evolution of the CME velocity (blue), the flare emission intensity in the EUVI 304 {\AA} passband (black, a proxy of the flare soft X-ray flux), mean temperature (red), and total EM (cyan). The error in the velocity (marked by the vertical symbol size) mainly comes from the uncertainty of the height, which is taken as the standard deviation of 10 measurements.} \label{f4}
\end{figure*}

\textbf{Role of magnetic reconnection in the flare and CME.} The 2D temperature maps (Figure \ref{f4}a) and the time-distance plot of the temperature along the vertical slice (Figure \ref{f4}b) reveal the detailed temperature evolution of the heated region and flare loops. Before the onset of the flare, the CME cavity is actually hotter than the surrounding coronal plasma, supporting the recent argument that it is most likely a pre-existing hot magnetic flux rope\cite{cheng11,zhang12}. Due to the rise motion of the hot cavity, the plasma underneath the reconnection region is quickly heated, forming a hot region with an average temperature of $\sim$4 MK. In the hot region, some flare loops are discernible. As the reconnection continues, the hot region further ascends and extends, and its temperature rises to $>$5 MK. On the other hand, the originally formed hot flare loops gradually cool down to $\sim$2 MK, and become more distinctive in the temperature maps. Meanwhile, newer hot flare loops are formed, stacked over the cool ones. The above process continues until the reconnection stops.

We deduce in detail the evolution of the emission intensity, DEM-weighted temperature, total emission measure (EM) of the flare region, and the velocity of the CME (Figure \ref{f4}c). The CME velocity is mostly synchronous with the intensity increase and plasma heating of the flare but precedes the total EM by tens of minutes. From $\sim$04:00 to 05:30 UT, the CME velocity, the flare intensity and the temperature rapidly increase. From $\sim$05:30 to 08:00 UT, the CME is still being accelerated, but the intensity and temperature of the flare reach a maximum with only some minor fluctuations. Afterwards, the intensity and temperature start to decrease; the CME velocity also decreases to a nearly constant value, most likely as a result of the interaction with the background solar wind\cite{gopal06}. Based on the velocity of the inflows and downflows, a lower limit on the reconnection rate (the inflow Alfv{\'e}n Mach number) is estimated to be 0.001--0.03, which is able to produce the observed weak solar flare and the long-accelerating slow CME.

\textbf{Discussion}

We have reconstructed the 3D magnetic topology of the fast reconnection in a solar eruption and quantify the properties of the reconnection and its role in the flare and CME. The method of direct imaging overcomes the disadvantage of magnetic field extrapolations based on non-linear force-free field modeling when studying a dynamic process\cite{cheng14}. The excellent observations provide much needed elucidation of the physical processes involved in a flare/CME in a 3D configuration. In a traditional 2D flare model\cite{sturrock66}, the initial magnetic configuration consists of two sets of oppositely directed field lines. A current sheet formed between them is an essential ingredient for flare occurrence. Once the reconnection begins, magnetic energy is released to produce an enhanced flare emission, and post-flare loops  are formed, mapping to two flare ribbons on the chromosphere. The discovery of a pre-existing flux rope makes a crucial addition to the standard paradigm\cite{zhang12}. It suggests that the reconnection is associated with the eruption of the flux rope. Although the flux rope is a 3D structure, observations made thus far of the reconnection are mostly restricted to 2D, in which the flux rope often appears as a hot plasma blob when viewed along the axis, and the reconnection site underneath the blob is apparently manifested as a thin and long sheet\cite{cheng11}. In a 3D case, the magnetic topology becomes much more complex and there are different regimes of magnetic reconnection\cite{pontin12}. The event studied here does reveal some new features. First, the reconnection site is more likely a separator or quasi-separator. The fact that the two sets of loops that are obviously non-coplanar are approaching each other does imply the presence of a separator or quasi-separator between them. Second, in the 2D case, reconnection forms an isolated closed field (a section of a flux rope) above, in addition to a flare loop below; while in the 3D case, reconnection supplies poloidal flux to the flux rope whose two ends are still anchored on the solar photosphere. As the reconnection proceeds, more and more poloidal flux is added to the flux rope, further accelerating the CME and in turn strengthening the flare emission. Our results are consistent with and lend observational support to models of flux rope induced solar eruptions\cite{shibata95}. The primary trigger of the eruption may be torus or kink instability of a pre-existing flux rope\cite{kliem06}, while magnetic reconnection, which occurs at a newly formed separator\cite{longcope05,parnell10} or quasi-separator\cite{aulanier05}, releases free magnetic energy and helps accelerate the eruption.

%\cite{pontin12}
\textbf{Methods}

\textbf{3D reconstruction and visualization.} Using the Interactive Data Language (IDL) program ``scc$\_$measure.pro'' in the Solar SoftWare (SSW) package, we reconstruct the 3D coordinates of the magnetic loops. This routine allows us to select a point in, for example, the AIA image. A line representing the line-of-sight from the AIA perspective is then displayed in the image from other perspectives, such as EUVI. According to the emission characteristics, we identify the same point at this line. 3D coordinates of the selected point (heliographic longitude, latitude, and radial distance in solar radii) are then determined. With the same manipulation, the 3D coordinates of all the points along the magnetic loops are derived. For each loop, the reconstruction is repeated 10 times, the most optimal one of which is chosen as the result, thus ensuring the accuracy of the reconstruction. In order to trace the evolution of the magnetic loops, we keep their footpoints fixed. The 3D visualization is realized by the software Paraview. 

\textbf{DEM reconstruction.} The differential emission measure (DEM) is recovered from six AIA passbands including 94 {\AA} (Fe X, $\sim$1.1 MK; Fe XVIII, $\sim$7.1 MK), 131 {\AA} (Fe VIII, $\sim$0.4 MK; Fe XXI, $\sim$11 MK), 171 {\AA} (Fe IX, $\sim$0.6 MK), 193 {\AA} (Fe XII, $\sim$1.6 MK), 211 {\AA} (Fe XIV, $\sim$2.0 MK), and 335 {\AA} (Fe XVI, $\sim$2.5 MK) through the regularized inversion method\cite{hannah12}. The observed flux $F_{i}$ for each passband can be written as:
 
 \begin{equation}
 {F_{i}}  =  \int DEM(T) R_i (T) \mathrm{d}T + \delta F_{i},
 \end{equation}
where $R_{i}(T)$ is the temperature response function of passband $i$, $DEM(T)$ indicates the plasma DEM in the corona, and $\delta F_{i}$ is the error of the observational intensity for passband $i$. The temperature range in the inversion is chosen as 5.5$\leq$ log${T}\leq$ 7.5. With the derived DEM, the DEM-weighted (mean) temperature and the total EM are calculated as:
 \begin{equation}
 T_{\mathrm{mean}} = \frac{ \int_{T_{\mathrm{min}}}^{T{\mathrm{max}}} DEM(T) T \mathrm{d}T} {\int_{T {\mathrm{min}}}^{T{\mathrm{max}}} DEM(T) \mathrm{d}T}
 \end{equation}
 and
 \begin{equation}
 EM= \int_{T {\mathrm{min}}}^{T {\mathrm{max}}} DEM(T) dT.\\
 \end{equation}
The temperature range of integration is set to be 5.7$\leq$ log${T}\leq$ 7.1, within which the EM solutions are well constrained as shown in Supplementary Figure 1. Finally, with the mean temperature at each pixel, the 2D temperature maps are constructed.

\textbf{Enhancement of EUV images.} To display the fine structures of the EUV images, we enhance the contrast by the routine ``aia$\_$rfilter.pro'' in SSW. This program first sums five images and divides the summed image into a number of rings. Each ring is then scaled to the difference of the maximum brightness and the minimum one. The final images are obtained by performing the Sobel edge enhancement taking advantage of the IDL program ``sobel.pro".

\textbf{3D Magnetic Field Extrapolation and Singularity Calculation.} With a potential field model, we extrapolate the 3D global magnetic field structure using the Helioseismic and Magnetic Imager\cite{schou12} daily updated synoptic maps of the radial magnetic field component on 2012 January 26 as the lower boundary. We further calculate the locations of all the null points and separators in the hemisphere containing the reconnection region. However, we can not find any null points or separators in the source region of the reconnection event. It implies that the null point or separator responsible for the reconnection event is probably formed during the initial stages of the eruption. Note that, since the magnetic data were measured one day before the event, a possible evolution of the photospheric magnetic field could change this conclusion.

\textbf {Uncertainty Analysis.} The errors of the DEM-weighted temperature and total EM depend on the errors of the DEM results, which come mainly from uncertainties in the temperature response functions of AIA including non-ionization equilibrium effects, non-thermal populations of electrons, modifications of dielectronic recombination rates, radiative transfer effects, and even the unknown filling factor of the plasma\cite{judge10}. Three representative DEM curves are shown in supplementary Figure 1b, from which one can find that the DEM solutions are well constrained in the temperature range 5.7$\leq$ log${T}\leq$ 7.1. In order to ensure the accuracy of the regularized inversion method, we also calculate the DEM with the forward fitting method\cite{cheng12} and find that the two results are very similar.

A possible deviation in the 3D reconstruction of magnetic topology mainly comes from the uncertainty in identifying the same feature from two different perspectives. However, this does not affect qualitatively the global 3D topology. The uncertainty in displaying the heated region is mostly from the assumption that the filling factor is 1 and the 3D temperature distribution in the hot region is of cylindrical symmetry with the cross section of the cylinder corresponding to the DEM-weighted 2D temperature map.

\textbf {Code availability.} The codes ``scc$\_$measure.pro'', ``aia$\_$rfilter.pro'', and ``sobel.pro" used in the above analysis are available at the website http://www.lmsal.com/solarsoft/.
\vspace{0.06\textwidth}

\textbf{Acknowledgements:} SDO is a mission of NASA's Living With a Star Program, STEREO is the third mission in NASA's Solar Terrestrial Probes Program, and SOHO is a mission of international cooperation between ESA and NASA. X.C., J.Q.S., M.D.D., Y.G., P.F.C., and C.F. are supported by NSFC through grants 11303016, 11373023, 11203014, and 11025314, and by NKBRSF through grants 2011CB811402 and 2014CB744203. C.E.P. and S.J.E. are supported by the UK STFC. J.Z. is supported by US NSF AGS-1249270 and AGS-1156120.

\textbf{Author Contributions} 
X.C. and J.Q.S. analyzed the observational data and contributed equally to the work. X.C. initiated the study. M.D.D. supervised the project and led the discussions. Y.G. joined part of the data analysis. C.E.P. and S.J.E. performed the magnetic field extrapolation. X.C. wrote the first draft. M.D.D. and E.R.P. made major revisions of the manuscript. Other authors discussed the results and commented on the manuscript.

\textbf{Additional information}
Correspondence and requests for information should be
addressed to X. Cheng (xincheng@nju.edu.cn) \& M.D. Ding (dmd@nju.edu.cn).

\textbf{Competing Financial Interests}
The authors declare no competing financial interests.

%\bibliographystyle{apj}
%\bibliography{reference}

\clearpage
%%%-----------------Figures-----------------------

\end{document}